\documentclass[12pt,3p,preprint]{elsarticle}

\usepackage{graphicx}
\usepackage{bm} 
\usepackage{epsfig}
\usepackage[latin1]{inputenc}
\usepackage{float,amsmath}
\usepackage{amssymb}

\def\lsim{\mathrel{\rlap{\lower4pt\hbox{$\sim$}}
    \raise1pt\hbox{$<$}}}                
\def\gsim{\mathrel{\rlap{\lower4pt\hbox{$\sim$}}
    \raise1pt\hbox{$>$}}}                
\newcommand{\beq}{\begin{equation}}
\newcommand{\eeq}{\end{equation}}
\newcommand{\bqa}{\begin{eqnarray}}
\newcommand{\eqa}{\end{eqnarray}}
\newcommand{\phard}{p_{\rm hard}}
\newcommand{\nn}{\nonumber}

\begin{document}


\begin{frontmatter}

\title{Dissipative Dynamics of Highly Anisotropic Systems}

\author[itp,fias]{Mauricio Martinez}
\address[itp]{
Institut f\"{u}r Theoretische Physik \\
Goethe-Universit\"{a}t Frankfurt \\
Max-von-Laue Strasse 1\\
D-60438, Frankfurt am Main, Germany
}
\author[gettysburg,fias]{Michael Strickland}
\address[gettysburg]{
  Physics Department, Gettysburg College\\
  Gettysburg, PA 17325 United States
}
\address[fias]{
Frankfurt Institute for Advanced Studies\\
Ruth-Moufang-Strasse 1\\
D-60438, Frankfurt am Main, Germany
}

\begin{abstract}

In this paper we present a method to improve the description of 0+1 dimensional boost invariant
dissipative dynamics in the presence of large momentum-space anisotropies.
We do this by reorganizing the canonical hydrodynamic 
expansion of the distribution function around a momentum-space
anisotropic ansatz rather than an isotropic equilibrium one.  At leading order the result 
obtained is two coupled ordinary differential equations for the momentum-space anisotropy and typical
momentum of the degrees of freedom.  We show that this framework can 
reproduce both the ideal hydrodynamic and free streaming limits.
Additionally, we demonstrate that when linearized the differential equations reduce to 
2nd order Israel-Stewart viscous hydrodynamics.  Finally, we make quantitative comparisons of
the evolution of the pressure anisotropy within our approach and
2nd order viscous hydrodynamics in both the strong and weak coupling limits.

\end{abstract}


\begin{keyword}
Boost Invariant Dynamics, Anisotropic Plasma, Non-equilibrium Evolution, Viscous Hydrodynamics.
\end{keyword}

\end{frontmatter}


\section{Introduction}
\label{sec:introduction}

The remarkable success of relativistic hydrodynamics to describe the anisotropic flow of
matter created during non-central heavy ion collisions 
\cite{Huovinen:2001cy, Hirano:2002ds,Tannenbaum:2006ch, Kolb:2003dz}
has sparked a great deal interest in the systematic derivation and application of viscous 
hydrodynamical equations for relativistic systems 
\cite{Dusling:2007gi,Luzum:2008cw,Song:2008hj,Heinz:2009xj,PeraltaRamos:2009kg,PeraltaRamos:2010je}.  
Here we demonstrate a method for deriving hydrodynamic-like equations for systems which have large 
momentum-space anisotropies.  Such large momentum-space anisotropies can arise naturally due to the rapid
longitudinal expansion of the matter created during relativistic heavy ion collisions.
These anisotropies can become so large that the shear becomes as large as the isotropic 
pressure, signaling the breakdown of the expansion of the energy-momentum tensor
in terms of shear corrections.  In fact, in 2nd order viscous hydrodynamics large momentum-space
anisotropies can even cause the longitudinal pressure predicted to become negative \cite{Martinez:2009mf}.

Here we present a way to circumvent such problems and derive evolution equations
which can describe the dynamics of systems with potentially large momentum-space anisotropies.  
The method is  to change the usual hydrodynamic expansion of the one-particle distribution functions
\beq
f(\tau,{\bf x},{\bf p}) = f_{\rm eq}(|{\bf p}|,T(\tau)) + \delta f_1 + \delta f_2 + \cdots \; ,
\eeq
which is expansion about an isotropic equilibrium state $f_{\rm eq}(|{\bf p}|,T(\tau))$, to one
in which the expansion point itself can contain momentum-space anisotropies
\beq
f(\tau,{\bf x},{\bf p}) = f_{\rm aniso}({\bf p},\phard(\tau),\xi(\tau)) + \delta f_1^\prime + \delta f_2^\prime + \cdots \; ,
\label{eq:newexp}
\eeq
where in both equations $\tau$ is proper time.
In Eq.~(\ref{eq:newexp}) $\xi$ is a parameter which measures the amount of momentum-space anisotropy and
$\phard$ is a non-equilibrium momentum scale which can be identified with the
temperature of the system only in the limit of isotropic equilibrium.
We will introduce an ansatz for $f_{\rm aniso}$  which approximates
the system via ellipsoidal equal occupation number surfaces as opposed to the spherical
ones associated with the isotropic equilibrium distribution function \cite{Romatschke:2003ms}.  
For highly anisotropic systems  one immediately expects that the corrections $\delta f_n^\prime$
will have smaller magnitude than the isotropic corrections $\delta f_n$ due to the fact that 
momentum-space anisotropy is built into the leading order of the expansion in the reorganized
approach.  One can use different prescriptions for an anisotropic basis.  Here we have in mind
the use of spheroidal harmonics \cite{abramowitz_stegun64} for which our ellipsoidal ansatz is the leading order term. 

Using this as a starting point, we will derive leading order 0+1 dimensional boost-invariant evolution equations for $f_{\rm aniso}$ by taking moments of the Boltzmann equation.\footnote{By leading order, we mean neglecting the corrections from the $\delta f_n^\prime$  terms in Eq.~(\ref{eq:newexp}).}
The result will be two coupled ordinary differential
equations which describe the evolution of the system's momentum-space anisotropy and typical hard
momentum scale.  We demonstrate that at leading order the new expansion can equally well describe
both the ideal hydrodynamic and free streaming limits within a unified framework.  In addition, we 
demonstrate analytically that when the associated coupled nonlinear differential equations are expanded
to leading order in the anisotropy parameter, they reduce identically to the 2nd order viscous hydrodynamic
equations of Israel and Stewart \cite{Israel:1976tn,Israel:1979wp,Muronga:2001zk,Muronga:2003ta}.  
In the general case we solve
the coupled nonlinear equations numerically allowing us to describe both the early-time high-anisotropy
dynamics and late-time near-equilibrium dynamics using a single formalism which gives us as
output the time-evolving distribution function.  This distribution function can then be used as input
for other calculations.

\section{Deriving boost invariant dynamics from transport theory}
\label{sec:deriving}

We consider a highly populated system formed by massless particles. We describe this
system by means of the Boltzmann equation for the one-particle distribution function 
$f(t,{\bf p},{\bf x})$ in the lab frame. 
Additionally, we assume that the system is boost invariant and expands only along the longitudinal 
(beam line) axis.  Boost invariance implies that that the longitudinal velocity of the system is constant along
lines of constant spatial rapidity such that $v_z = z/t$.  
The assumption that the system only expands in the longitudinal direction is reliable if the
transverse size of the system is sufficiently large ($\sim 1.2\,A^{1/3}$ for central nucleus-nucleus 
collisions) that the effects of transverse dynamics can be ignored to first approximation.  If this is the case
then one can assume a homogeneous distribution in the transverse directions and set $v_x = v_y = 0$.  With these assumptions one
can show that the dynamics reduces to 0+1 dimensional evolution, i.e., only derivatives
with respect to proper time remain.  Here we will show how this occurs by starting in lab coordinates
and transforming the Boltzmann equation to comoving coordinates.  Under the assumption of homogeneity in the transverse 
direction one has in lab coordinates
\beq
p^t \partial_t f(t,z,{\bf p}) + p^z \partial_z f(t,z,{\bf p})= -{\cal C}[f(t,z,{\bf p})] \, ,
\label{eq:boltzmann}
\eeq
where ${\cal C}[f(t,z,{\bf p})]$ is the collisional kernel which is a functional of the distribution function. Its functional form
depends on the type of interactions between the particles and, in general, is difficult to evaluate exactly even for 
simple theories.   To make analytic progress in this paper we will study the case that the
collisional kernel is given by the relaxation time approximation
\beq
{\cal C}[f(t,z,{\bf p})] = p_\mu u^\mu \, \Gamma \, \left[ f(t,z,{\bf p}) - f_{\rm eq}(t,z,|{\bf p}|,T(
\tau)) \right]\,,
\label{rel-time}
\eeq
where $f_{\rm eq}$ is the local equilibrium distribution function and $\Gamma$ is the relaxation rate. 
The temperature in the local rest frame $T(\tau)$ is determined dynamically by requiring 
${\cal E}_{\rm eq}(\tau)={\cal E}_{\rm non\hbox{-}eq}(\tau)$, 
that guarantees energy conservation within this approximation~\cite{Baym:1984np}
and $T(\tau)$ is assumed to only depend on the proper time $\tau = \sqrt{t^2-z^2}$ in accordance with 
the assumption of boost invariance.
Although here we consider the relaxation time approximation, we note that the general method 
can be applied to collisional kernels derived directly from quantum field 
theoretical methods.  We postpone such studies to future publications.

\subsection{Ansatz for the distribution function and equation of state}
\label{subsec:RSansatz}

In order to have a tractable approach that will allow us to describe systems that may become
highly anisotropic in momentum space we consider a general ansatz for our anisotropic distribution 
function which was first introduced by Romatschke and Strickland (RS)~\cite{Romatschke:2003ms}. Within 
this approach the leading order anisotropic distribution function in the local rest frame
can be obtained from an arbitrary isotropic distribution  function ($f_{\rm iso}$) by squeezing 
($\xi>0$) or stretching ($\xi<0$) $f_{\rm iso}$ along one direction in momentum space
\beq
f(\tau,{\bf x},{\bf p})=f_{RS}({\bf p},\xi(\tau),\phard(\tau))=f_{\rm iso}\bigl([{\bf p^2}+\xi(\tau)({\bf p\cdot \hat{n}})^2]/p^2_{\rm hard}(\tau)\bigr) \; ,
\label{eq:distansatz}
\eeq
where $p_{\rm hard}$ is related to the average momentum in the partonic distribution function, $\hat{\bf n}$ is the direction of the 
anisotropy,\footnote{Hereafter, we use $\hat{\bf n}=\hat{\bf e}_z$, where $\hat{\bf e}_z$ is a unit vector along the longitudinal 
direction.} and $-1 < \xi < \infty$ is a parameter that reflects the strength and type of anisotropy. The anisotropy parameter 
$\xi$ is related to the average longitudinal and transverse momentum of the constituents via the 
relation~\cite{Mauricio:2007vz,Martinez:2008di}
\beq
\xi=\frac{\langle p_T^2\rangle}{2\langle p_L^2\rangle} - 1 \; .
\label{anisoparam}
\eeq
The system is locally isotropic when $\xi=0$ but this does not imply that it is in local thermal equilibrium unless $f_{\rm iso}$ is
an equilibrium distribution function. 

The energy-momentum tensor in the local rest (comoving) frame with coordinates 
$x^\mu = (\tau=\sqrt{t^2-z^2},x,y,\varsigma={\rm arctanh}(z/t))$ is given by
$T^{\mu\nu}= (2\pi)^{-3}\,\int d^3{\bf p}/p^0\, p^\mu p^\nu f(\tau,{\bf x},{\bf p})$.
In the comoving frame the energy-momentum tensor is diagonal and using the RS ansatz (\ref{eq:distansatz}) its components 
are~\cite{Martinez:2009ry}
\begin{subequations}
\label{momentsanisotropic}
\begin{align}
\label{energyaniso}
{\cal E}(p_{\rm hard},\xi) &= T^{\tau\tau} \;= \frac{1}{2}\left(\frac{1}{1+\xi}
+\frac{\arctan\sqrt{\xi}}{\sqrt{\xi}} \right) {\cal E}_{\rm iso}(p_{\rm hard}) \; , \\ \nonumber
&\equiv{\cal R}(\xi)\,{\cal E}_{\rm iso}(p_{\rm hard})\, ,\\
\label{transpressaniso}
{\cal P}_T(p_{\rm hard},\xi) &= \frac{1}{2}\left( T^{xx} + T^{yy}\right) 
= \frac{3}{2 \xi} 
\left( \frac{1+(\xi^2-1){\cal R}(\xi)}{\xi + 1}\right)
 {\cal P}_{\rm iso}(p_{\rm hard}) \, , 
\\ \nonumber
&\equiv{\cal R}_{\rm T}(\xi){\cal P}_{\rm iso}(p_{\rm hard})\, , \\
\label{longpressaniso}
{\cal P}_L(p_{\rm hard},\xi) &= - T^{\varsigma}_\varsigma= \frac{3}{\xi} 
\left( \frac{(\xi+1){\cal R}(\xi)-1}{\xi+1}\right) {\cal P}_{\rm iso}(p_{\rm hard}) \; ,\\ \nonumber
&\equiv {\cal R}_{\rm L}(\xi){\cal P}_{\rm iso}(p_{\rm hard})\, ,
\end{align}
\end{subequations}
where ${\cal P}_{\rm iso}(p_{\rm hard})$ and ${\cal E}_{\rm iso}(p_{\rm hard})$ are the isotropic 
pressure and energy density, respectively. In general, one cannot 
identify immediately ${\cal P}_{\rm iso}(p_{\rm hard})$ and ${\cal E}_{\rm iso}(p_{\rm hard})$ 
with their equilibrium counterparts, unless one implements the Landau matching 
conditions~\cite{Martinez:2009ry}.~\footnote{In general the Landau matching conditions are a way to connect 
		non-equilibrium evolution to near-equilibrium hydrodynamic evolution.  In practice, this corresponds
		to matching the components of the energy-momentum tensor and matching collective velocities \cite{landaulifschitz}.}
For a conformal system the tracelessness of the stress-energy tensor $T^\mu_\mu$=0 implies 
${\cal E} = 2{\cal P}_T + {\cal P}_L$. This condition is satisfied by Eqs.~(\ref{momentsanisotropic}) for any anisotropic 
distribution function~(\ref{eq:distansatz}) since for an isotropic conformal state ${\cal E}_{\rm iso}=3{\cal P}_{\rm iso}$.

In addition to the components of the energy-momentum tensor, one can also calculate the entropy density in the comoving frame 
following the standard definition from kinetic theory~\cite{GLW}. For the case of the RS ansatz~(\ref{eq:distansatz}) one obtains~\cite{Martinez:2009ry}
\bqa
{\cal S}&=&-\int \frac{d^3 {\bf p}}{(2\pi)^3}\,f(\tau,{\bf x},{\bf p})\bigl\{\log\bigl[f(\tau,{\bf x},{\bf p})\bigr]-1\bigr\}\,, \nonumber\\
&=&\frac{{\cal S}_{\rm iso} (p_{\rm hard})}{\sqrt{1+\xi}} \,, 
\label{entropydens}
\eqa 
where ${\cal S}_{\rm iso} (p_{\rm hard})$ is the isotropic entropy density.

An additional property of the RS ansatz~(\ref{eq:distansatz}) is that in the case of 0+1 dimensional boost invariant 
expansion, it is possible to relate the anisotropy parameter $\xi$ and the shear used in viscous hydrodynamical
treatments by 
matching the values of the longitudinal pressure ${\cal P}_{\rm L}$. To see this, consider the 
longitudinal pressure for 0+1 dimensional viscous hydrodynamics with an ideal equation of state
\beq
\label{visclongpres}
{\cal P}_{\rm L}(\tau)={\cal P}_{\rm eq}(T(\tau))-\Pi(\tau)=\frac{{\cal E}_{\rm eq}(T(\tau))}{3}
\left(1-\frac{3\Pi(\tau)}{{\cal E}_{\rm eq}(T(\tau))}\right)\,,
\eeq
where $\Pi \equiv \Pi^\varsigma_\varsigma$ is the $\varsigma\varsigma$ component 
of the fluid shear tensor.
The local temperature $T(\tau)$ and hard momentum scale $\phard$ contained in the RS ansatz~(\ref{eq:distansatz}) are 
related through the Landau matching condition
for the energy density, ${\cal E}_{\rm eq}(\tau)={\cal E}_{\rm non\hbox{-}eq}(\tau)$, which implies that at any time $\tau$ we can relate $T$ and $\phard$ via
$T={\cal R}^{1/4}(\xi)\phard$~\cite{Martinez:2009ry}. By using this relation, Eq.~(\ref{longpressaniso}) can be rewritten as
\bqa
\label{longpressmatch}
{\cal P}_{\rm L}(\tau)&=&{\cal R}_{\rm L}(\xi(\tau)){\cal P}_{\rm T}^{\rm iso}(\phard(\tau))=
\frac{{\cal R}_{\rm L}(\xi(\tau))}{{\cal R}(\xi(\tau))}{\cal P}_{\rm T}^{\rm eq}(T(\tau))\nn\\
&=&\frac{{\cal E}_{\rm eq}(T(\tau))}{3}\,\frac{{\cal R}_{\rm L}(\xi(\tau))}{{\cal R}(\xi(\tau))}\,.
\eqa
Equating Eqs.~(\ref{visclongpres}) and (\ref{longpressmatch})  we find
\beq
\label{piallorders}
\Pi(\tau)=\frac{{\cal E}_{\rm eq}(T(\tau))}{3}\left[1-\frac{{\cal R}_{\rm L}(\xi(\tau))}{{\cal R}(\xi(\tau))}\right]\,,
\eeq
which is valid to all orders in $\xi$. In the case of small departures from equilibrium the system is
very nearly isotropic with $\Pi\ll {\cal E}$ and one can use a small $\xi$ expansion to 
obtain~\cite{Martinez:2009ry}

 \beq
\label{smallxiviscous}
\frac{\Pi}{{\cal E}_{\rm eq}}= \frac{8}{45}\,\xi+{\cal O}(\xi^2)\,,
\eeq
where ${\cal E}_{\rm eq}$ is the equilibrium energy density. If one uses the Navier-Stokes value of the shear tensor 
$\Pi_{\rm NS}=4\eta/(3\tau)$ and the ideal equation of state in the last relation, we obtain~\cite{Asakawa:2006jn}
\beq
\xi_{\rm NS} = \frac{10}{T \tau}\frac{\eta}{{\cal S}}+{\cal O}(\Pi^2_{\rm NS})\,,
\eeq
where $\eta$ is the shear viscosity and ${\cal S}$ is the equilibrium entropy of the system.
Therefore, deviations from equilibrium which are canonically inferred from the evolution of the shear viscous tensor can be also 
studied using the RS distribution function~(\ref{eq:distansatz}) by analyzing the evolution of the anisotropy parameter $\xi$. 

We point out that $p_{\rm hard}$ and $\xi$ are time-dependent variables whose evolution equations are unknown a priori. In this work, 
we derive and solve such equations by taking the first two moments of the Boltzmann equation within the relaxation time 
approximation. This procedure allows us to study their time evolution and obtain information 
about the transition between early-time non-equilibrium dynamics and late-time viscous hydrodynamical behavior. 

\subsection{Boltzmann equation in comoving coordinations}
\label{subsec:Boltzeqn}

It is convenient to switch to comoving coordinates
\bqa
t&=&\tau \cosh\varsigma \, , \nn \\
z&=&\tau \sinh\varsigma \, ,  
\label{com-coord}
\eqa
and introduce momentum-space rapidity
\bqa
p^0 &=& p_T \cosh y , \, \nn \\
p^z &=& p_T \sinh y \, .
\label{yrapidity}
\eqa
In this coordinate system the metric $g_{\mu\nu}= {\rm diag}\,(1,-1,-1,-\tau^2)$ and the RS ansatz (\ref{eq:distansatz}) 
is $f_{\rm RS}({\bf p},\xi,\phard)=f_{\rm iso}\!\left(p_T^2[1+(1+\xi)\sinh^2 (y-\varsigma)]/\phard^2\right)$. After changing variables, the 
Boltzmann equation~(\ref{eq:boltzmann}) can be written as
\bqa
&&\hspace{-1.5cm}\left(p_T \cosh (y-\varsigma)\frac{\partial}{\partial\tau}+
\frac{p_T\,\sinh (y-\varsigma)}{\tau}\,\frac{\partial}{\partial\varsigma}\right) f_{\rm RS}({\bf p},\xi,\phard) = \nn\\
&&\hspace{-0.2cm}-\Gamma\, p_T \cosh (y-\varsigma)\bigl[ f_{\rm RS}({\bf p},\xi,\phard)- f_{\rm eq}({\bf p}, T(\tau))\bigr]\,.
\label{eq:RSBoltz}
\eqa
Note that in the above equation one should evaluate the last term on the right side using the constraint
$T(\tau)=R^{1/4}(\xi)\phard$ which is required by energy conservation~\cite{Baym:1984np,Martinez:2009ry}.

The usual way to study non-equilibrium deviations within a kinetic theory approach is to take moments of the Boltzmann equation
~\cite{Grad}. For instance, by using the $14$ moments ansatz~\cite{Israel:1976tn,Israel:1979wp} it is possible to determine the 
equation of motion for the shear viscous tensor by taking the 2nd moment of the Boltzmann equation. In this work, we just need to 
determine the equations of motion for $\xi$ and $\phard$ by considering the zeroth and first moments of the Boltzmann equation
~(\ref{eq:RSBoltz}). We demonstrate this in the next two sections. 

\subsubsection{Zeroth moment of the Boltzmann equation}
\label{subsubsec:zerothmoment}

The zeroth moment of the Boltzmann equation with respect to the single particle energy reduces
to $\partial_\mu N^\mu$ on the left hand side~\cite{GLW}; however, in the relaxation time approximation the
zeroth moment of the collisional kernel on the right hand side is non-vanishing.  Since in any realistic non-equilibrium quantum field theory there are generically 
number non-conserving processes we consider this to be a useful feature of the
relaxation time approximation.~\footnote{We 
note that if one were considering a number-conserving theory, one can enforce $\partial_\mu N^\mu=0$ in the relaxation
time approximation by introducing a particle fugacity into the ansatz for the one particle distribution function.  One would then need
to take one further moment of the Boltzmann equation to determine the dynamical equations.}

To derive the first equation, let us first rewrite the left hand side (LHS) 
of Eq.~(\ref{eq:RSBoltz}) explicitly using the RS ansatz 
\beq
{\rm  LHS} = E\, \partial_\omega f_{\rm RS} (\omega)\left[\frac{p_z^2}{\phard^2}
\left(\partial_\tau \xi - \frac{2(1+\xi)}{\tau}\right) - \frac{2\omega}{\phard}\,\partial_\tau \phard\right]\,,
\eeq
where $\omega \equiv \left[p_T^2+(1+\xi)p_z^2\right]/\phard^2$ and we have rewritten the result in terms of the local energy and longitudinal momentum defined in the fluid
cell via $E= p_T\cosh (y-\varsigma)$ and $p_z=p_T\sinh (y-\varsigma)$, respectively. By using this expression together with the right hand side of Eq.~(\ref{eq:RSBoltz}) we can take the zeroth moment by integrating both sides with the momentum-space measure 
$(2\pi)^{-3} \int d^3{\bf p}/(2p^0)$. This results in
\bqa
\label{AAA}
&&\hspace{-2cm}\frac{\bigl\langle p_z^2\,\partial_\omega f_{\rm RS}(\omega)\bigr\rangle}{\phard^2} 
\left(\partial_\tau \xi - \frac{2(1+\xi)}{\tau}\right) -\frac{2\bigl\langle \omega\,
\partial_\omega f_{\rm RS}(\omega)\bigr\rangle }{\phard}
\partial_\tau \phard \nn\\
&& \hspace{2cm}=-\Gamma \biggl[\bigl\langle f_{\rm RS}({\bf p},\xi,\phard) \bigr\rangle-
\bigl\langle f_{\rm eq}\bigl({\bf p}, T(\tau)) \bigr\rangle\biggr]
\eqa
where $\langle a \rangle\equiv (2\pi)^{-3}\int d^3{\bf p}\, a({\bf p}) $. These integrals can be calculated analytically 
using the general form of the RS ansatz~(\ref{eq:distansatz})
\begin{subequations}
\begin{align}
\bigl\langle p_z^2\,\partial_\omega f_{\rm RS}(\omega)\bigr\rangle &= 
-\frac{1}{2}\frac{\phard^2}{(1+\xi)^{3/2}}\,n_{\rm iso} (\phard) \, , \\
\bigl\langle \omega \,\partial_\omega f_{\rm RS}(\omega)\bigr\rangle &= 
-\frac{3}{2}\,\frac{n_{\rm iso} (\phard) }{(1+\xi)^{1/2}} \, , \\
\bigl\langle f_{\rm RS}({\bf p},\xi,\phard) \bigr\rangle &= \frac{n_{\rm iso}(\phard)}{(1+\xi)^{1/2}} \, , \\
\bigl\langle f_{\rm eq}\bigl({\bf p}, T(\tau)={\cal R}^{1/4}(\xi)\phard\bigr) \bigr\rangle&= 
{\cal R}^{3/4}(\xi)\,n_{\rm iso} (\phard)  \, , 
\end{align}
\end{subequations}
where in the last line we have used the Landau matching condition $T(\tau)={\cal R}^{1/4}(\xi)\phard$. After replacing these 
expressions in Eq.~(\ref{AAA}), we obtain
\beq
\frac{1}{1+\xi} \partial_\tau \xi- \frac{2}{\tau} - \frac{6}{\phard} \partial_\tau \phard = 
2 \Gamma \left[ 1 - {\cal R}^{3/4}(\xi) \sqrt{1+\xi} \right] \, ,
\label{eq:zerothmoment}
\eeq
which is our first coupled differential equation for $\xi$ and $\phard$.

\subsubsection{First moment of the Boltzmann equation}
\label{subsubsec:firstmoment}
In this section we derive the second coupled differential equation for $\xi$ and $\phard$. Taking the first moment of the 
Boltzmann equation is equivalent to energy-momentum conservation 
$\partial_\mu T^{\mu\nu}=0$~\cite{GLW}.  We enforce energy conservation
by requiring that ${\cal E}_{\rm eq}(\tau)={\cal E}_{\rm non\hbox{-}eq}(\tau)$ which
results in $T(\tau)=R^{1/4}(\xi)\phard$.  Momentum conservation is guaranteed for
any distribution function which is parity-symmetric in momentum space. 
In the comoving frame, energy conservation implies that the equation of motion 
for the energy density is
\beq
\label{eq:energydens}
\frac{\partial {\cal E (\tau)}}{\partial \tau}=-\frac{{\cal E} (\tau)+{\cal P}_L(\tau)}{\tau} \, ,
\eeq
where $\cal{E} $ and ${\cal P}_L$ are the energy density and longitudinal pressure. Using the components
of the energy momentum tensor given by the RS ansatz (Eqs.~(\ref{energyaniso}) and~(\ref{longpressaniso})) we obtain 
\beq
\frac{{\cal R}'(\xi)}{{\cal R}(\xi)} \partial_\tau \xi + \frac{4}{\phard} \partial_\tau \phard = 
\frac{1}{\tau} \left[ \frac{1}{\xi(1+\xi){\cal R}(\xi)} - \frac{1}{\xi} - 1 \right] \, ,
\label{eq:firstmoment}
\eeq
where
\beq
{\cal R}'(\xi)=\partial_\xi {\cal R}(\xi)=\frac{1}{4}\biggl(\frac{1-\xi}{\xi(1+\xi)^2}-\frac{\text{arctan}
\sqrt{\xi}}{\xi^{3/2}}\biggr)\,.
\eeq
Eq.~(\ref{eq:firstmoment}) is the second coupled ordinary differential equation for $\xi$ and $\phard$. 

\subsection{Combined Differential Equations}

It is possible to further simplify Eqs.~(\ref{eq:zerothmoment}) and~(\ref{eq:firstmoment}) to obtain
\begin{subequations}
\label{eq:mix}
\begin{align}
\frac{1}{1+\xi}\partial_\tau\xi &= \frac{2}{\tau} - 4 \, \Gamma \, {\cal R}(\xi) \, \frac{{\cal R}^{3/4}(\xi)\sqrt{1+\xi}-1}{2 {\cal R}(\xi) + 3 (1+\xi) {\cal R}'(\xi)} \, , \\
\frac{1}{1+\xi}\frac{1}{\phard} \partial_\tau\phard &= \Gamma \, {\cal R}'(\xi) \, \frac{{\cal R}^{3/4}(\xi)\sqrt{1+\xi}-1}{2 {\cal R}(\xi) + 3 (1+\xi) {\cal R}'(\xi)} \, .
\end{align}
\end{subequations}
In the results section we show that the solution of these equations allows us to describe the transition between early-time longitudinal free streaming and late time hydrodynamical behavior. 
Note that one should expect that the solutions of these equations 
reduce to either free streaming expansion or ideal hydrodynamical behavior when $\Gamma\to0$ or $\Gamma\to\infty$, respectively. In 
the next section we show this by taking the asymptotic limits of these equations.

\subsection{Asymptotic limits}
\label{subsec:asymtlimit}

One way of checking self-consistency in our approach is to see if it is possible to obtain from Eqs.~(\ref{eq:mix}) the 
two extreme cases of expansion: (1) a system which undergoes 0+1 dimensional longitudinal free streaming expansion and (2) a system which is ``instantaneously'' thermal and isotropic and undergoes 0+1 
dimensional ideal hydrodynamical expansion. In this section we show that Eqs.~(\ref{eq:mix}) contain
both of these limits naturally.

\subsubsection{Free Streaming Limit}
\label{subsubsec:freestr}

When the particles do not interact the interaction rate $\Gamma$ vanishes exactly. In order to show that the free streaming limit is recovered in our approach, we set $\Gamma=0$ in Eqs.~(\ref{eq:mix}).  One obtains immediately
\begin{subequations}
 \label{eq:freestreaminglimit}
\begin{align}
\partial_\tau \xi &= \frac{2}{\tau}(1+\xi) \, , \\
\partial_\tau \phard &= 0 \, .
\end{align}
\end{subequations}
The solutions of these equations are 
\begin{subequations}
\label{eq:freestreamingsols}
 \begin{align}
  \xi(\tau)&=(1+\xi_0)\left(\frac{\tau}{\tau_0}\right)^2-1 \,,\\
  \phard&=p_0\,,
 \end{align}
\end{subequations}
where $\xi_0$ and $p_0$ are the initial values of $\xi$ and $\phard$ at $\tau=\tau_0$. This corresponds precisely to the free streaming 
solution to the Boltzmann equation~\cite{Baym:1984np,Martinez:2008di,Martinez:2009ry,Kapusta:1992uy}.

\subsubsection{Ideal hydrodynamical expansion}
\label{subsubsec:idealhydro}

In the case that the system is isotropic at all times we have $\xi(\tau) = 0$ and therefore
$\partial_\tau \xi =0$.  This occurs in the limit that $\Gamma \to \infty$.
From Eq.~(\ref{eq:zerothmoment}), we can expand the right hand side
\bqa
\lim_{\xi \rightarrow 0} 2\Gamma \left[ 1 - R^{3/4}(\xi) \sqrt{1+\xi} \right]&=& 
2\Gamma \left[1-(1+{\cal O}(\xi^2))\right]\nn\,,\\&=&0\,,
\eqa
so that in this limit Eq.~(\ref{eq:zerothmoment}) gives us
\beq
\frac{1}{\phard} \partial_\tau \phard = - \frac{1}{3 \tau} \, .
\label{eq:idealhydrolimit-1}
\eeq
We can also consider the same limit in Eq.~(\ref{eq:firstmoment}). First, we expand the right hand side
\beq
\lim_{\xi \rightarrow 0} \frac{1}{\xi(1+\xi){\cal R}(\xi)} - \frac{1}{\xi} - 1= -\frac{4}{3}+ {\cal O} (\xi)\,,
\eeq
so that Eq.~(\ref{eq:firstmoment}) reduces to 
\beq
\frac{1}{\phard} \partial_\tau \phard = - \frac{1}{3 \tau} \, .
\label{eq:idealhydrolimit-2}
\eeq
Eqs.~(\ref{eq:idealhydrolimit-1}) and (\ref{eq:idealhydrolimit-2}) are exactly the same and the solution is 
$\phard(\tau)=\phard(\tau_0)\,(\tau_0/\tau)^{1/3}$ which corresponds to the solution for the ideal 0+1 boost invariant 
expansion~\cite{Bjorken:1982qr}.\footnote{We point out that in the limit $\xi \to 0$, the local equilibrium temperature $T(\tau)$
is exactly the same as the typical momentum of the particles in the system $\phard$ since ${\cal R}(0)=1$.} 

\subsection{Positivity of entropy divergence}

One can prove analytically that the solutions to Eqs.~(\ref{eq:mix}) have positive entropy divergence.  Formally
the requirement is stated as $\partial_\mu {\cal S}^\mu \geq 0$ where ${\cal S}^\mu$ is the entropy current.
In the 0+1 dimensional case this simplifies to the requirement $\partial_\tau (\tau {\cal S}) \geq 0$ where ${\cal S}$
is the anistropic entropy defined in Eq.~(\ref{entropydens}).  Using Eq.~(\ref{entropydens}) we first use the chain rule
to obtain
\beq
\frac{\partial_\tau (\tau {\cal S})}{ {\cal S}} = 1 - \frac{\tau}{2(1+\xi)} \partial_\tau \xi + \frac{3 \tau}{\phard} \partial_\tau\phard \, .
\eeq
Using Eqs.~(\ref{eq:mix}) and simplifying we find
\beq
\frac{\partial_\tau (\tau {\cal S})}{ {\cal S}}  = \tau \Gamma \left[ {\cal R}^{3/4}(\xi)\sqrt{1+\xi} - 1 \right] \, .
\eeq 
Since ${\cal R}^{3/4}(\xi) \sqrt{1+\xi} \geq 1$ for all $\xi$ in $-1 < \xi < \infty$ and all other quantities are positive, 
we therefore always have $\partial_\tau (\tau {\cal S}) \geq 0$.  Note that in the ideal hydrodynamical limit ($\xi=0$)
and the free streaming limit ($\Gamma=0$) the lower bound of the inequality is reached with $\partial_\tau(\tau {\cal S}) = 0$
as is expected.

\subsection{2nd order viscous hydrodynamics from RS ansatz}
\label{subsec:2ndvisc}

In Sect.~\ref{subsec:RSansatz} we showed that in the near-equilibrium limit one can relate the viscous shear tensor $\Pi$ with the 
anisotropy parameter $\xi$ (Eq.~(\ref{smallxiviscous})). As a consequence, in our approach, the evolution of dissipative 
corrections to the ideal fluid is equivalent to the evolution of the anisotropy parameter $\xi$. To see this more in detail, 
in this section we derive the equations of motion for 2nd order 0+1 boost invariant 
viscous hydrodynamics from the RS ansatz~(\ref{eq:distansatz}) by linearizing the
previously obtained coupled nonlinear differential equations.  

We begin by using Eq.~(\ref{eq:energydens}) for the energy density, Eq.~(\ref{longpressmatch}) for the matching between $\xi$ and $\Pi$, and assume
an ideal equation of state.\footnote{During the derivation presented in this section, it is necessary to implement the Landau matching condition 
which relates the local temperature and the average momentum of the particles $\phard$, i.e. $T(\tau)={\cal R}^{1/4}(\xi(\tau))\phard(\tau)$.}  The result is
\bqa
\partial_\tau {\cal E}= -\frac{{\cal E}}{\tau}\left(1+\frac{{\cal R}_{\rm L}(\xi)}{3{\cal R}(\xi)}\right)\,,
\eqa
which upon using~(\ref{piallorders}) gives
\bqa
\label{eq:energyvisceq}
\partial_\tau {\cal E}&=&-\frac{{\cal E}+{\cal P}}{\tau}+\frac{\Pi}{\tau} \; .
\eqa
This expression is the well-known equation for the energy
density when the system has shear viscous corrections in a 0+1 dimensional boost-invariant expansion
~\cite{Israel:1976tn,Israel:1979wp,Muronga:2001zk}.

To obtain the equation of motion for the shear viscous tensor we use Eq.~(\ref{eq:zerothmoment})
derived in Sect.~\ref{subsubsec:zerothmoment}. First, let us rewrite some terms which appear in this equation by using the linear relation between $\xi$ and $\Pi$
\beq
\partial_\tau \xi = \frac{45}{8}\left(\frac{\partial_\tau \Pi}{{\cal E}}-
\frac{\Pi}{{\cal E}}\frac{\partial_\tau {\cal E}}{{\cal E}}\right)\,,
\eeq
where ${\cal E}$ is understood to be the isotropic equilibrium energy density in this section. Next, by implementing the Landau 
matching condition for the local temperature and $\phard$ we obtain
\beq
\frac{\partial_\tau \phard}{\phard}=\frac{1}{4}\left(\frac{\partial_\tau {\cal E}}{\cal E}-\frac{{\cal R}'(\xi)}{{\cal R}(\xi)}
\partial_\tau\xi\right) \, .
\eeq
Plugging the last two expressions into Eq.~(\ref{eq:zerothmoment}), using Eqs.~(\ref{smallxiviscous}) and (\ref{eq:energyvisceq}), and 
expanding to the lowest non-vanishing order in $\xi$ one obtains
\beq
\label{eq:visctenseqin}
\partial_\tau \Pi +\frac{4}{3}\frac{\Pi}{\tau}-\frac{16}{45}\frac{{\cal E}}{\tau}= -\frac{1}{2}\Gamma\,\Pi\,.
\eeq
To finally connect with second order viscous hydro, we identify\footnote{This is analogous to the identifications of the 
unknown coefficients $\beta_0$, $\beta_1$ and $\beta_2$ which appear in the 14 Grad's ansatz
~\cite{Israel:1976tn,Israel:1979wp,Muronga:2001zk,Muronga:2003ta,Denicol:2010xn,El:2009vj}.}
\begin{subequations}
\bqa
\Gamma &=& \frac{2}{\tau_\pi}\,, \\ 
\tau_\pi &=& \frac{5}{4}\frac{\eta}{\cal P}\,,
\eqa
\label{eq:hydromatch}
\end{subequations}
so Eq.~(\ref{eq:visctenseqin}) can be written finally as
\beq
\label{eq:visctenseq}
\partial_\tau\Pi=-\frac{\Pi}{\tau_\pi}+\frac{4}{3}\frac{\eta}{\tau_\pi\tau}-\frac{4}{3}\frac{\Pi}{\tau}\,,
\eeq
where $\tau_\pi$ and $\eta$ are the shear relaxation time and shear viscosity, respectively.
This last expression is precisely the 0+1 dimensional 2nd order viscous hydrodynamical equation for $\Pi$
~\cite{Israel:1976tn,Israel:1979wp,Muronga:2001zk}. 
Therefore, the physics of 2nd order viscous hydrodynamics can be recovered from our ansatz in the small $\xi$ limit. 

\section{Results}
\label{sec:results}

\begin{figure}[t]
\begin{center}
\includegraphics[width=13.5cm]{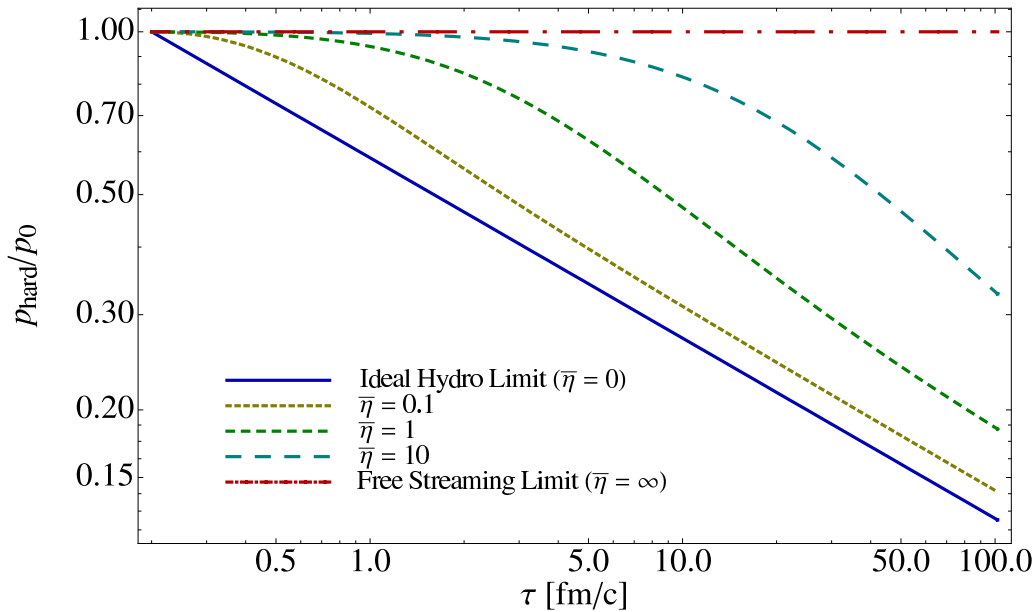}
\end{center}
\vspace{-6mm}
\caption{
Hard momentum scale $\phard$ as a function of proper time for five different values of $\bar\eta$:  the ideal hydrodynamic limit ($\bar\eta=0$), $\bar\eta = 0.1$, $\bar\eta = 1$, $\bar\eta = 10$, and the free streaming limit ($\bar\eta = \infty$).
}
\label{fig:p0vsetabar}
\end{figure}

\begin{figure}[t]
\begin{center}
\includegraphics[width=13.5cm]{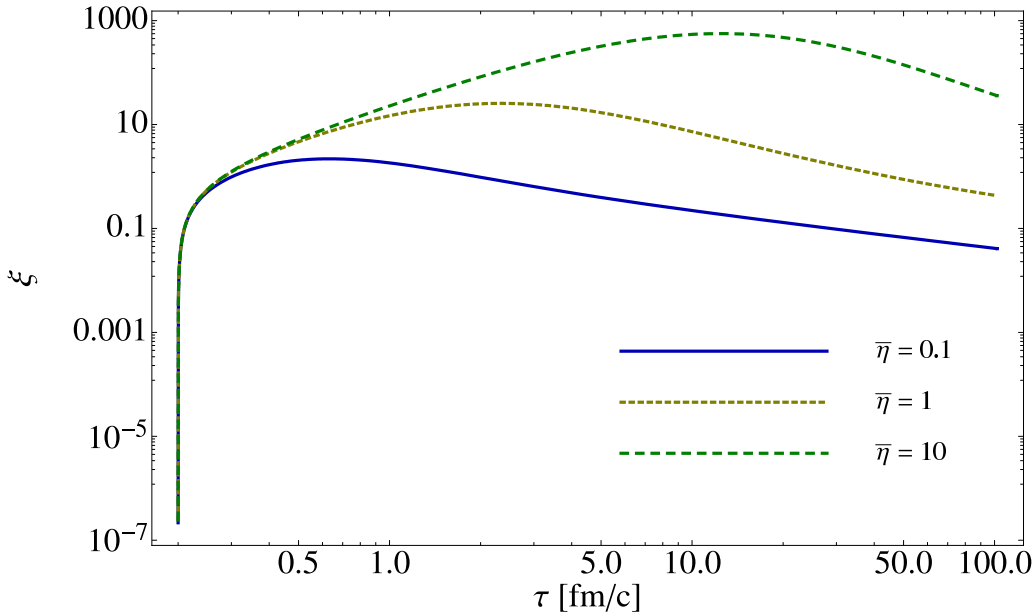}
\end{center}
\vspace{-6mm}
\caption{
Anisotropy parameter $\xi$ as a function of proper time for three different values of $\bar\eta$:  $\bar\eta = 0.1$, $\bar\eta = 1$, and $\bar\eta = 10$.
}
\label{fig:xivsetabar}
\end{figure}

\begin{figure}[t]
\begin{center}
\includegraphics[width=13.5cm]{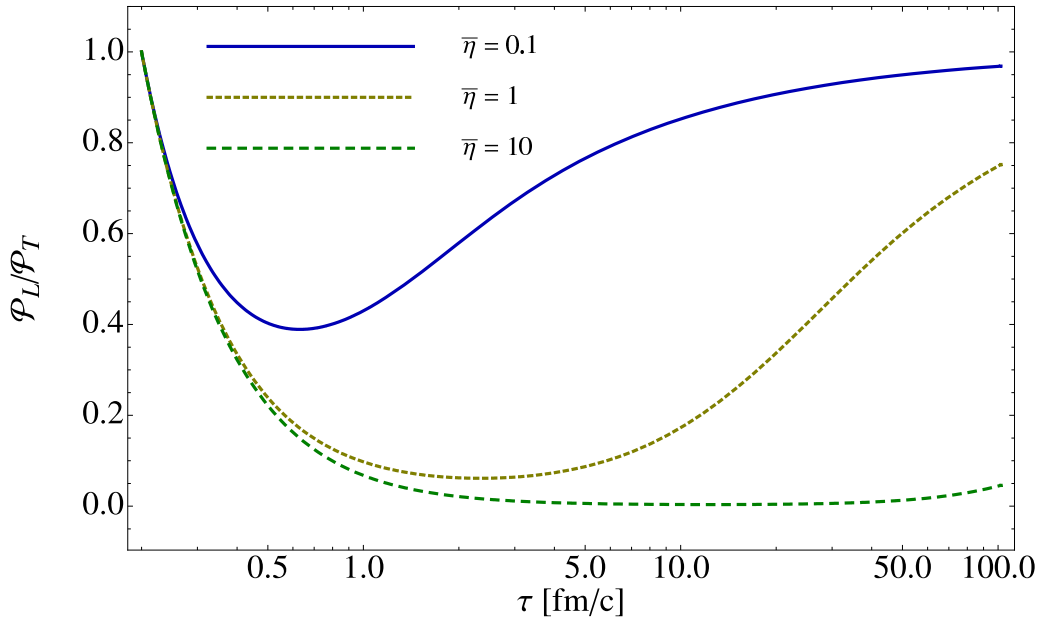}
\end{center}
\vspace{-6mm}
\caption{
Ratio of longitudinal [Eq.~(\ref{longpressaniso})] and transverse [Eq.~(\ref{transpressaniso})] pressures as a function of proper time for three different values of $\bar\eta$:  $\bar\eta = 0.1$, $\bar\eta = 1$, and $\bar\eta = 10$.
}
\label{fig:ploptvsetabar}
\end{figure}

\begin{figure}[t]
\begin{center}
\includegraphics[width=13.5cm]{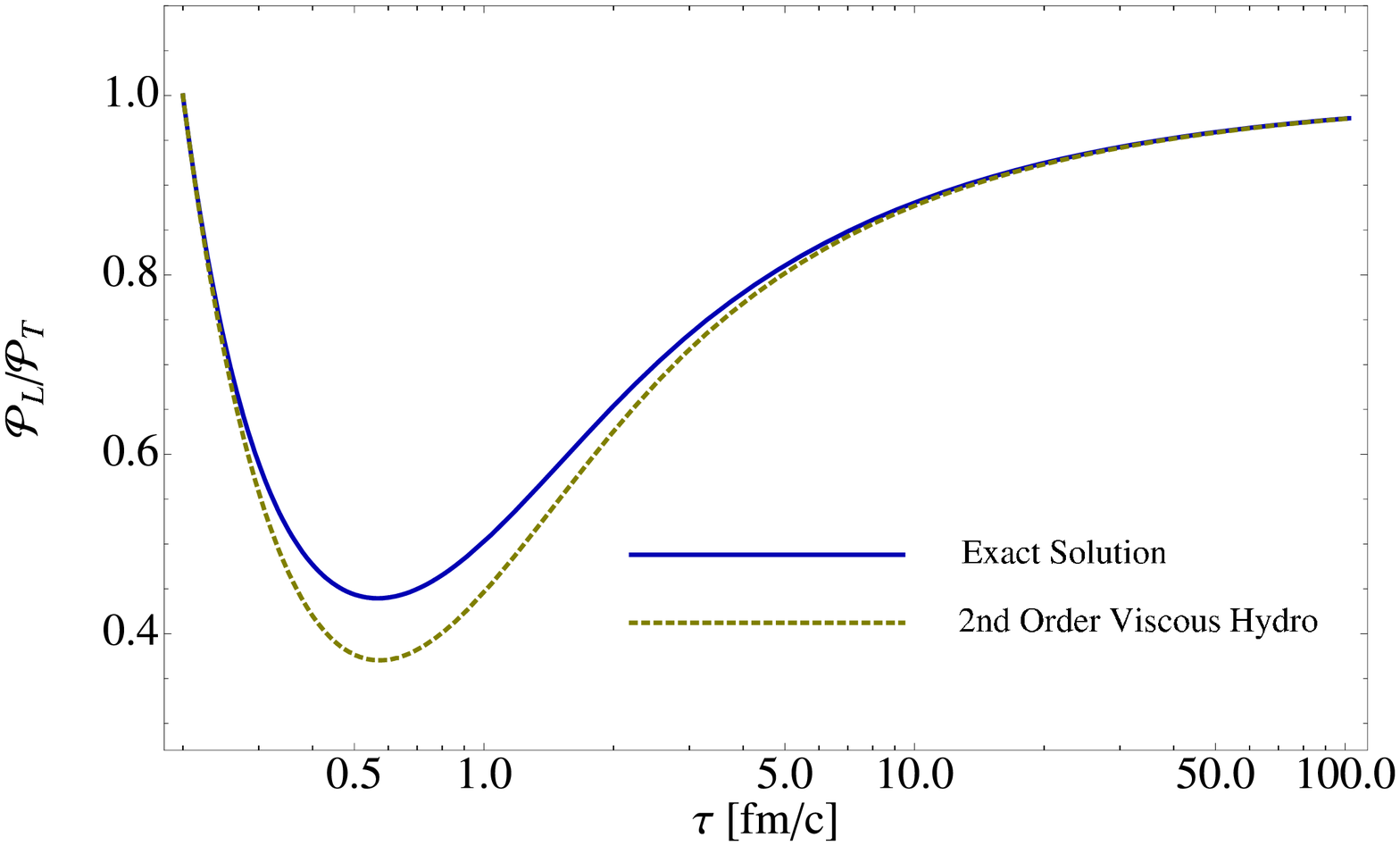}
\end{center}
\vspace{-6mm}
\caption{
Ratio of longitudinal [Eq.~(\ref{longpressaniso})] and transverse [Eq.~(\ref{transpressaniso})] pressures as a function of proper time for $\bar\eta = 1/(4\pi)$.  Solid line is numerical solution to Eqs.~(\ref{eq:mix}) and dashed line is numerical solution of 2nd order viscous hydrodynamics given by equations (\ref{eq:energyvisceq}) and (\ref{eq:visctenseq}).
}
\label{fig:ploptstrongcoupling}
\end{figure}

\begin{figure}[t]
\begin{center}
\includegraphics[width=13.5cm]{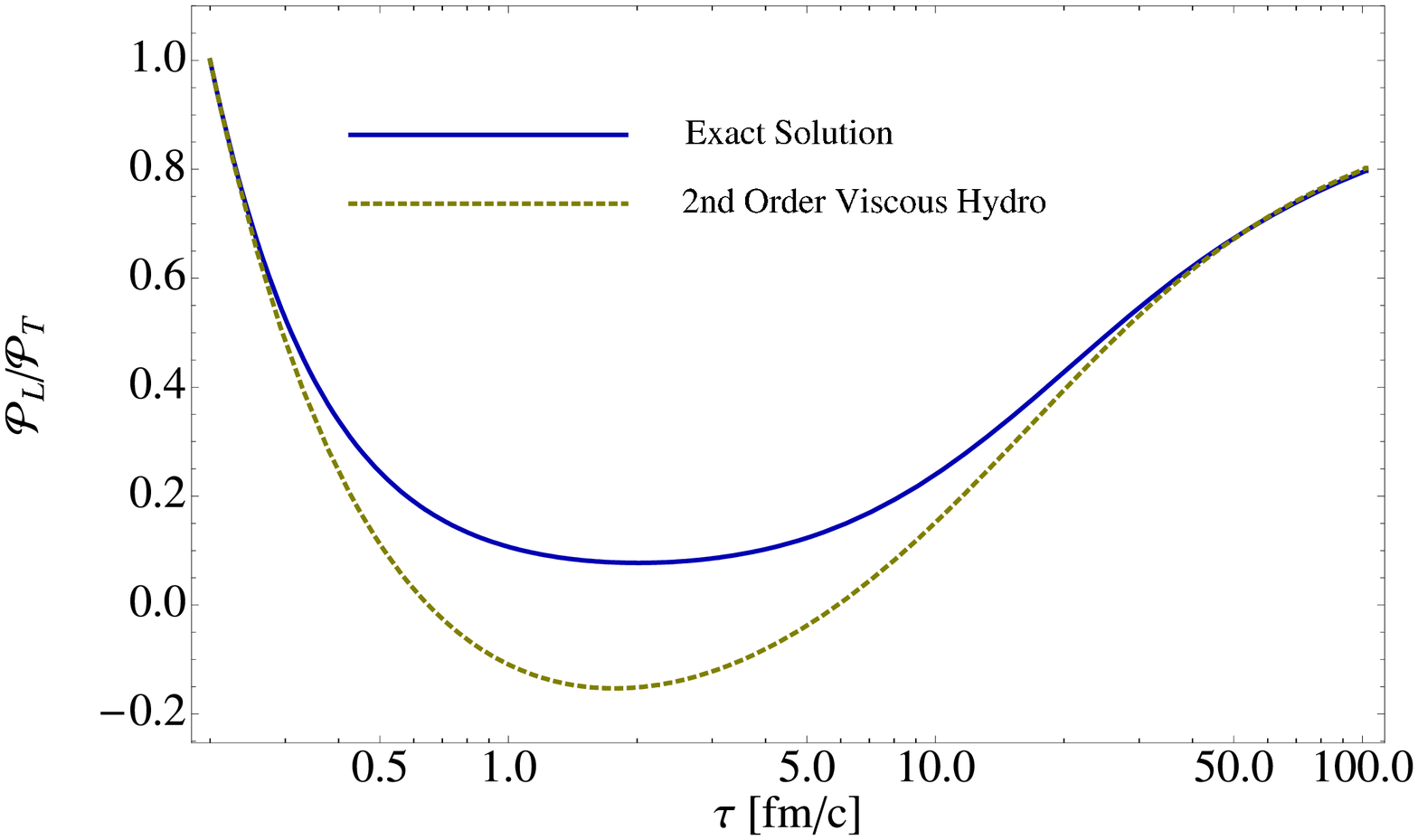}
\end{center}
\vspace{-6mm}
\caption{
Ratio of longitudinal [Eq.~(\ref{longpressaniso})] and transverse [Eq.~(\ref{transpressaniso})] pressures as a function of proper time for $\bar\eta = 10/(4\pi)$.  Solid line is numerical solution to Eqs.~(\ref{eq:mix}) and dashed line is numerical solution of 2nd order viscous hydrodynamics given by equations (\ref{eq:energyvisceq}) and (\ref{eq:visctenseq}).
}
\label{fig:ploptweakcoupling}
\end{figure}

In this section we present the results of numerically integrating the coupled nonlinear differential equations 
using the relation between the relaxation rate $\Gamma$ and the ratio of shear viscosity to equilibrium entropy $\bar\eta \equiv \eta/{\cal S}$ which results from Eqs.~(\ref{eq:hydromatch}), namely
\beq
\Gamma = \frac{2T(\tau)}{5\bar\eta} = \frac{2{\cal R}^{1/4}(\xi)\phard}{5\bar\eta} \, ,
\label{eq:gammamatch}
\eeq
which should be used for $\Gamma$ in Eqs.~(\ref{eq:mix}).
In all plots shown in this paper we will use the following initial conditions:  $\tau_0$= 0.2 fm/c, $\phard(\tau=\tau_0)$ = 350 MeV, 
and $\xi(\tau=\tau_0) = 0$.  Note, however, that one can choose whatever initial conditions one likes in order to integrate
Eqs.~(\ref{eq:mix}). 

In Fig.~\ref{fig:p0vsetabar} we show the solutions obtained for the hard momentum scale $\phard$ as a function of proper time for five different values of $\bar\eta$:  the ideal hydrodynamic limit ($\bar\eta=0$), $\bar\eta = 0.1$, $\bar\eta = 1$, $\bar\eta = 10$, and the free streaming limit ($\bar\eta = \infty$).  As can be seen from this figure, the solutions to Eqs.~(\ref{eq:mix}) make a smooth transition from ideal hydrodynamic behavior to longitudinal free streaming as $\bar\eta$ goes from $0$ to $\infty$.  For finite values of $\bar\eta$ one sees that the solutions smoothly transition between early-time longitudinal free streaming and late-time viscous hydrodynamical behavior.

In Fig.~\ref{fig:xivsetabar} we show the corresponding solutions for the anisotropy parameter $\xi$ as a function of proper time for three different values of $\bar\eta$:  $\bar\eta = 0.1$, $\bar\eta = 1$, and $\bar\eta = 10$.  In the ideal hydrodynamic limit one has $\xi=0$ and in the free streaming limit one has $\xi = (1+\xi_0)(\tau/\tau_0)^2-1$.  The solutions presented in Fig.~\ref{fig:xivsetabar} show that at very early times the behavior of the system only depends weakly on $\bar\eta$ due to the rapid longitudinal expansion of the system.  In addition, in all cases, the anisotropy has a maximum which is obtained at later proper times as $\bar\eta$ is increased.

Knowledge of the proper-time dependence of $\phard$ and $\xi$ allows one to calculate all of the various thermodynamic functions defined in Eqs.~(\ref{energyaniso}), (\ref{transpressaniso}), (\ref{longpressaniso}), and (\ref{entropydens}).  For sake of brevity we do not show each of these individually but we note that the energy density smoothly transitions from early-time free streaming to late-time hydrodynamical expansion as one
would expect.  A quantity that is sensitive to momentum-space anisotropies is the ratio of the longitudinal to transverse pressures, ${\cal P}_L/{\cal P}_T$.  In the ideal hydrodynamic limit this is identically one and in the free streaming limit one obtains ${\cal P}_L/{\cal P}_T \propto 1/\tau^2$ at late times.  In Fig.~\ref{fig:ploptvsetabar} we show ${\cal P}_L/{\cal P}_T$ for three different values of $\bar\eta$:  $\bar\eta = 0.1$, $\bar\eta = 1$, and $\bar\eta = 10$.  One sees from this figure that as $\bar\eta$ is increased, the longitudinal pressure is decreased; however, at late times the system restores isotropy in momentum space for any finite $\bar\eta$.  It is important to note that even for the case $\bar\eta=10$ the longitudinal pressure of the system remains positive definite.

Having obtained the general numerical solutions to Eqs.~(\ref{eq:mix}) we can now make a quantitative comparison between
solution to the fully nonlinear differential equations (\ref{eq:mix}) and the linearized 2nd order viscous hydrodynamical equations given by
Eqs.~(\ref{eq:energyvisceq}) and (\ref{eq:visctenseq}).  In Figs.~\ref{fig:ploptstrongcoupling} and \ref{fig:ploptweakcoupling} we
compare the two approaches for typical strong and weak coupling values of $\bar\eta=1/(4\pi)$ and $\bar\eta=10/(4\pi)$, 
respectively.  As one can see from  Fig.~\ref{fig:ploptstrongcoupling} for $\bar\eta=1/(4\pi)$ the difference between the RS
ansatz solutions and 2nd order viscous hydro is at maximum approximately 15\%.  However, in the weak coupling case
of $\bar\eta=10/(4\pi)$ one finds that the correction can be greater than 100\%.  More importantly, one finds that in the weak
coupling case, for the chosen initial conditions, the ratio ${\cal P}_L/{\cal P}_T$ obtained from 2nd order viscous hydrodynamics becomes negative for $0.65 $ fm/c $\lsim \tau \lsim 5.9$ fm/c whereas the solution obtained
using the RS ansatz is positive definite during the entire evolution.  Quantitatively one should note that even in the strong coupling
case large momentum-space anisotropies can be developed with ${\rm min}({\cal P}_L/{\cal P}_T) \sim 0.45$ even for the exact solution.
The precise numbers, of course, depend on the choice of the initial conditions and proper time at which one begins integrating the differential equations.
When cast into dimensionless form the relevant quantity is $\tau_0 T$ with larger values of the dimensionless number corresponding
to generation of smaller momentum-space anisotropies.

\section{Conclusions and Outlook}
\label{sec:concl.}

In this paper we have shown that by starting from an ansatz which incorporates momentum-space anisotropies
from the beginning one can obtain solutions which smoothly connect the ideal hydrodynamic and free streaming
limits by varying $\bar\eta$ from zero to infinity.  We demonstrated that these limits are achieved analytically
and numerically.  In addition, we showed that when the equations are linearized the resulting linear differential equations reduce 
to the Israel-Stewart equations for viscous hydrodynamical evolution.  We then made a detailed comparison of
the prediction for the ratio of longitudinal to transverse pressure predicted by the RS ansatz and the Israel-Stewart
formalism.  We found that for typical strong coupling values of $\bar\eta$ the correction from 2nd order viscous
hydrodynamical evolution was on the order of 15\% and for a typical weak coupling value of $\bar\eta$ that
the correction was sizable and could be greater than 100\%.  

The conclusions reached here are based on one particular set of initial conditions, however, the pattern observed
here is generic.  At LHC energies the initial time of hydro evolution is expected to be less based on perturbative
estimates of the thermalization time; however, the initial temperature is also expected to be higher.  When cast into dimensionless
form what matters is the dimensionless combination $\tau_0 T_0$ for determining the magnitude of expected
longitudinal momentum-space anisotropies.  For the RHIC-like initial conditions presented 
here this combination is
$\tau_0 T_0 = 0.35$ whereas for LHC-like initial conditions one obtains $\tau_0 T_0 = 0.43$ assuming $\tau_0 = 0.1$ fm/c
and $T_0$ = 850 MeV.  Therefore, one expects slightly smaller momentum-space anisotropies to be generated for LHC initial conditions.

The method shown here can be used as input to calculate the dependence of high energy observables on the
time evolution of the anisotropy parameter and hard momentum scale.  There are now calculations of the anisotropic
photon rate \cite{Schenke:2006yp}, dilepton rate \cite{Mauricio:2007vz,Martinez:2008di,Martinez:2008mc}, 
and quarkonium binding energies \cite{Dumitru:2007hy,Dumitru:2009ni,Dumitru:2009fy} using the 
RS ansatz.  It would 
be interesting to apply the differential equations derived here to these phenomenological applications.

In addition, looking forward one can also extend the ansatz used here to allow for the inclusion of 
particle fugacities.  This can be done by evaluating another moment of the Boltzmann equation; however, one
will need to take care with the time evolution of the number density in this case if one wants to obtain solutions
which reach chemical equilibrium at late times \cite{El:2010mt}.  We postpone this development to a future paper.

\section*{Note Added}
During the final preparation of this manuscript another paper presenting a similar idea appeared by
Florkowski and Ryblewski \cite{Florkowski:2010cf}.  However, they obtain two differential
equations by requiring energy conservation and introducing an ansatz for an entropy source 
whereas we follow the method of taking moments of the Boltzmann equation directly.

\section*{Acknowledgments}
We thank A. El, G. Denicol, A. Dumitru, C. Greiner, P. Huovinen, H. Niemi, P. Romatschke, A. Rebhan, and Z. Xu 
for useful discussions.
M. Martinez and M. Strickland were supported by the Helmholtz International Center for FAIR 
Landesoffensive zur Entwicklung Wissenschaftlich-\"Okonomischer Exzellenz program.

\bibliographystyle{utphys}
\bibliography{anisodynamics}

\end{document}